# A POSSIBLE LOCAL COUNTERPART TO THE EXCESS POPULATION OF FAINT BLUE GALAXIES

Stacy S. McGaugh

Institute of Astronomy, University of Cambridge

Madingley Road, Cambridge CB3 0HA, UK

**OBSERVATIONS of galaxies to very faint magnitudes have revealed a population of blue galaxies at intermediate redshift[1-5]. These galaxies represent a significant excess over the expectation of standard cosmological models for reasonable amounts of evolution of the locally observed galaxy population. However, the surveys which define the local galaxy population are strongly biased against objects of low surface brightness[6-9]. Low surface brightness galaxies have properties very similar to those of the excess blue population[10,11], and recent work suggests that they are comparable in abundance to the more readily detected normal galaxies[9,12]. I show that the very deep surveys which reveal the excess population can easily detect low surface brightness galaxies to large redshifts, but that local surveys will miss them because they are not comparably sensitive. This suggests that the excess faint galaxies *are* low surface brightness galaxies. No alteration of standard cosmology is required, but it is necessary to reconsider the way in which the galaxy distribution function is specified.**

Great interest has been generated by the discrepancy between the number counts of galaxies in the $B$ and $I$ bands[1-5] with standard cosmological models ($0 < q_0 \leq 0.5$, little or no galaxy evolution for $z < 0.5$). Such models are consistent with the counts in the $K$ band[13,14] and the observed redshift distribution[14-16]. The problem is a substantial excess over model expectations in the number of galaxies with very blue spectral energy distributions.

Proposed solutions involve altering either cosmology or the galaxy population. The former changes the geometry of the universe[4,17] in order to reconcile the high counts with the low redshifts, while the latter supposes strong evolution in the numbers of galaxies[4,14], selective evolution in their luminosity[15,18], or an entirely new population[19]. The most common alteration of cosmology invokes a finite cosmological constant[17], but this may already be in conflict with limits imposed by the statistics of gravitational lenses[20]. Evolution in the numbers of galaxies provides a good description of the observations[4,14,21], but the high inferred rate of mergers at $z \sim 0.1$ is difficult to reconcile with the low current merger rate and limits on the mass accreted by spiral disks[22]. It also violates the observation that the faint blue galaxies are in relatively isolated environments[19,23] rather than the dense regions where merging can occur. Luminosity evolution must be selective[15] in order to match the apparent evolution in number density. In this hypothesis, most galaxy types evolve mildly, but the dwarf galaxies which are common and faint at the present epoch evolve strongly[15,18] so that there is a high density of bright objects at the appropriate redshift. The required amount of evolution is extreme, and is not observed to actually occur[24]. None of these explanations appear satisfactory, suggesting that the excess population of blue galaxies observed at moderate redshift may have a local counterpart which has remained unidentified.

There does exist locally a population of galaxies with physical properties similar to those of the faint blue galaxy population: low surface brightness disk galaxies. These are galaxies with central surface brightnesses well below what is considered[25] 'normal' for spirals ($\mu_0^B \gtrsim 23$ mag arcsec$^{-2}$ rather than $\mu_0^B \approx 21.6$ mag arcsec$^{-2}$). They have colors and luminosities[11,26] comparable to those of the excess population (Fig. 1), so invoking extreme forms of evolution is not necessary. Indeed, the star formation history inferred for low surface brightness galaxies[11] suggests that they should not change much in color and luminosity for $z < 0.4$. They remain blue because they form stars at a low, approximately constant rate, and have not had time to build up a substantial red giant branch since their relatively recent epoch of formation (very roughly, $z \sim 1$).

In addition, low surface brightness disks are observed to be weakly clustered[12,27], with a correlation amplitude lower than normal galaxies by about the same factor as the faint galaxy population[12,23]. Other observed properties of the excess faint population, such as star formation[14,16] and morphology[15], though difficult to quantify, appear to be consistent with those of low surface brightness galaxies[10]. This correspondence of physical properties makes it natural to identify the two populations with one another.

For this identification to be correct, the deep surveys which reveal the excess population must be more sensitive to low surface brightness galaxies than local surveys. At first this may seem counterintuitive, as cosmological $(1+z)^4$ dimming (due to the noneuclidean geometry of the universe) will decrease initially low surface brightness even further. This applies equally to all galaxies, however.

Surveys are usually said to be complete to some limiting flux, but strictly speaking this is only correct in the case of point sources. Since galaxies are resolved, it is also necessary to specify the isophotal threshold above which the flux is measured. The portion of a galaxy which is detected depends on the luminosity profile of that galaxy. The types of galaxies of interest here can be described as exponential disks with profiles of the form

$$\mu(r) = \mu_0 + 1.086\frac{r}{l}.$$

The two parameters $\mu_0$ and $l$ (central surface brightness and exponential scale length) characterize the surface brightness distribution of the disk of a galaxy and allow us to determine the ratio of observed flux ($F_{obs}$) to total flux ($F_{tot}$) which is actually measured by a survey. This ratio is

$$F_{obs}/F_{tot} = 1 - (1+n)e^{-n}$$

where $n$ is the number of scale lengths $l$ observed above the isophotal limit. This convenient analytical approximation, which will slightly underestimate the observed fraction of light from spirals with bulges, is nevertheless an important improvement over treating galaxies as point sources.

The isophotal aperture, in scale lengths $n$, of a survey conducted to a limiting surface brightness $\mu_l$ is given by

$$n = \frac{\mu_l - \mu_0 - 10log(1+z) - k(z)}{1.086}.$$

The terms involving $z$ account for $(1+z)^4$ surface brightness dimming and the redshifting of the spectral energy distribution (the $k$-correction[28]). The $k$-corrections depend on galaxy type as well as redshift, but are negligibly small for the blue galaxies of interest here.

The great sensitivity of the deep surveys more than overcomes $(1+z)^4$ dimming and low surface brightness galaxies can be seen to high redshifts because essentially all the flux is detected (Fig. 2). For the typical observed redshift ($z \sim 0.4$), $F_{obs}/F_{tot} > 0.8$ for $\mu_0^B < 24$ mag arcsec$^{-2}$, and even modest sized galaxies exceed the flux limit. Thus, detecting low surface brightness galaxies at intermediate redshift is not a problem.

It is rather more of a problem locally. Normal spirals[25] can be detected to distances $\sim$ 3 times as great as low surface brightness disks of the same size, and hence sample a volume $\sim 27$ times larger. This causes the numbers of low surface brightness galaxies to be seriously underestimated. Even when selected, isophotal measures systematically underestimate the flux of low surface brightness systems. This effect is severe; for $\mu_0^B = 23$ mag arcsec$^{-2}$, $F_{obs}/F_{tot} \approx 0.45$, which drops to $F_{obs}/F_{tot} \approx 0.12$ for $\mu_0^B = 24$ mag arcsec$^{-2}$. This causes systematic errors in the derived luminosity function and number counts in the observed sense[30,31] — the luminosity function will be too flat, and the number counts too steep. The steep slope of the number counts at relatively bright magnitudes[31] is incompatible with any reasonable amount of galaxy evolution for the short time since $z \sim 0.1$, but is expected from the use of isophotal magnitudes when the distribution of galaxy surface brightnesses is not a delta function.

The extended nature of galaxies means that it is necessary to consider the bivariate distribution of luminosity *and* surface brightness rather than the luminosity function alone. The bivariate distribution is not well quantified in the field[32], but in clusters the surface brightness distribution is observed to be flat[33,34] rather than sharply peaked[25]. To illustrate the effects on the luminosity function derived by various surveys, the expression for $F_{obs}/F_{tot}$ can be combined with assumed forms of the bivariate distribution (Fig. 3). The luminosity function is usually taken to be of the form[35]

$$\phi(L) = \phi^* \left(\frac{L}{L^*}\right)^\alpha e^{-L/L^*}$$

where $\phi^*$ describes the density of galaxies, $L^*$ is a characteristic luminosity brighter than which galaxies are rare ($L_B^* \sim 10^{10} h^{-2} L_\odot$), and $\alpha$ is the asymptotic slope of the faint end. These parameters can be seriously mistaken by standard procedures which implicitly assume that $F_{obs}/F_{tot}$ is a constant. In particular, local surveys are prone to underestimating the numbers of objects with $L < \frac{1}{2} L^*$ where the discrepancy with faint counts occurs.

Specifically, it is possible to find bivariate distributions which reproduce the locally determined luminosity function and predict an apparent excess in the numbers of objects seen in deep surveys. Indeed, the problem is degenerate, as there are many forms of the bivariate distribution which can reproduce the observations. If, as seems likely, the true bivariate distribution is intermediate between the extreme cases illustrated in Fig. 3, so that there is an average trend of luminosity with surface brightness with broad scatter, then both the density of galaxies $\phi^*$ and the slope $|\alpha|$ (which describes the frequency of faint galaxies) will be underestimated locally. This may be sufficient to explain the entire discrepancy in the faint galaxy number counts.

Identification of the faint galaxy excess with low surface brightness galaxies is actually a conservative hypothesis, as no drastic alterations to cosmology or radical forms of galaxy evolution are required. Large numbers of low surface brightness galaxies are already indicated locally by their high surface number density[9] and the simple model which matches their spatial distribution[12]. Since isophotal measures are inadequate when a distribution of surface brightnesses is present, it is necessary to operate from the premise of measuring the bivariate distribution of galaxies rather than leaping directly to the luminosity function.

While galaxy evolution inevitably occurs, it will be impossible to measure from number counts, much less derive cosmological parameters, until the bivariate distribution is quantified. Indeed, correcting for redshift effects requires detailed knowledge of the spectral energy distribution ($SED$), so interpreting data on galaxies at significant redshift requires knowledge of the trivariate distribution function $\Psi(L, \mu_0, SED)$. If low surface brightness galaxies maintain their blue colors into the ultraviolet, they may actually be the most prominent objects at $z \sim 0.4$.

Galaxy distributions of the sort suggested here can alleviate many outstanding problems in extragalactic astronomy which, like the faint blue galaxies, require more objects than provided by the apparent local luminosity function. Examples include the large number of Ly$\alpha$ absorption systems along the line of sight to QSOs, the baryonic mass missing relative to the otherwise successful predictions of primordial nucleosynthesis theory, and the high measured value of the extragalactic background radiation. Theories of cosmic structure formation should predict the bivariate distribution, perhaps removing the need for ad hoc mechanisms to flatten the predicted luminosity function to match local observations. More fundamentally, we need to ask what is meant by frequently used qualitative terms like 'normal' galaxies. By this, do we mean a type of galaxy which is common (like faint main sequence stars), or one which merely happens to be optically prominent (like red giants)?

ACKNOWLEDGEMENTS. I am grateful to many colleagues for conversations which led to this work, in particular M. Ress, B. E. J. Pagel, and S. D. M. White. I acknowledge support from a SERC postdoctoral fellowship.

# FIGURE CAPTIONS

FIG. 1  The $V - I$ color vs. $B$ luminosity for a sample of faint[15] (open circles) and low surface brightness[11] (filled squares) galaxies. The faint galaxy sample shows the expected number of red galaxies, but an excess in the number of galaxies with blue colours ($V - I \sim 1$). The faint galaxy data have been shifted to the rest frame following ref. 15 and assuming $H_0 = 100 \mathrm{\,km\,s^{-1} Mpc^{-1}}$. The excess faint blue galaxies occupy the same region of this diagram as do low surface brightness galaxies.

FIG. 2  The fraction of the total flux ($F_{tot}$) observed ($F_{obs}$) within isophotal apertures as a function of redshift for galaxies with exponential profiles. The specific case of the selection characteristics of the deep survey of ref. 4 is shown in the upper right; that for a large area local survey[28,29] which is commonly taken to define the luminosity function[30] occupies the left part of the diagram. Dash-dotted lines of constant rest frame central surface brightness (in half magnitude steps labeled by $\mu_0^B$) decline because of cosmological $(1 + z)^4$ dimming. Solid lines delimit the distance to which galaxies of different sizes can be seen for the flux limits of the two surveys, including the effect of flux lost outside the aperture ($F_{obs}/F_{tot} < 1$). The solid lines are labeled by the size characteristic $l$, the exponential scale length.

The two surveys have very different selection characteristics. The deep survey detects nearly all of the flux of low surface brightness galaxies even at large redshifts, while the local survey detects only a fraction of their total flux. This causes low surface brightness galaxies to be missed locally; observing a population comparable to that detected by deep surveys would require a local survey with an isophotal limit more sensitive by $\sim 2.5$ magnitudes.

FIG. 3  The luminosity function derived from two forms of the bivariate galaxy distribution. a: a bivariate distribution in which luminosity and surface brightness are tightly correlated. b: no correlation between luminosity and surface brightness (see insets). These opposite forms of the bivariate distribution (the true form of which is not well known) illustrate the hazards posed by isophotal measures and surface brightness selection effects.

In (a), surface brightness falls off linearly with luminosity below $L^*$ (solid line in inset). This is not a realistic bivariate distribution, but neither is one in which the surface brightness distribution is a delta function (the dashed-triple dotted line in the inset, implicit in most determinations of the luminosity function which assume that $F_{obs}/F_{tot}$ is a constant). For an intrinsic luminosity function with a slope $\alpha = -1.5$ (the heavy solid line in the main figure), the dashed line will be observed by the deep survey of ref. 4 while the thin solid line will be observed by the local survey of ref. 30. These have been shifted in normalisation to agree with the observations at the bright end: the dotted line is the luminosity function actually derived in ref. 30 for disk type galaxies. The thin solid line provides as good a description of the observations as the dotted line, despite representing a very different intrinsic distribution. This occurs because the luminosities of low surface brightness galaxies are systematically underestimated, leading to an underestimate of the slope $|\alpha|$ and a translation in $L^*$ and $\phi^*$. The effect is more severe in the case of the local survey, causing an apparent excess in the numbers of galaxies observed by the deep survey (the difference between the dashed and thin solid line) even though both are drawn from the same intrinsic distribution.

In (b), galaxies fill the luminosity–surface brightness plane below a maximum[7] surface brightness without correlation (inset). This is a realistic bivariate distribution[32–34], though not well quantified in the field. In this case, the observed $\phi^*$ depends on the maximum surface brightness of the distribution and the limiting isophote of a survey. More sensitive surveys will probe further into the surface brightness distribution, again causing an apparent excess in the numbers of galaxies observed in deep relative to local surveys. There is no guarantee that there is a lower limit to the surface brightness distribution, hence the arrow on the intrinsic distribution. Detecting progressively fainter, higher redshift galaxies necessarily requires achieving progressively greater isophotal sensitivity, so a bivariate distribution of this sort predicts that the apparent $\phi^*$ should increase with $z$, as observed[21].

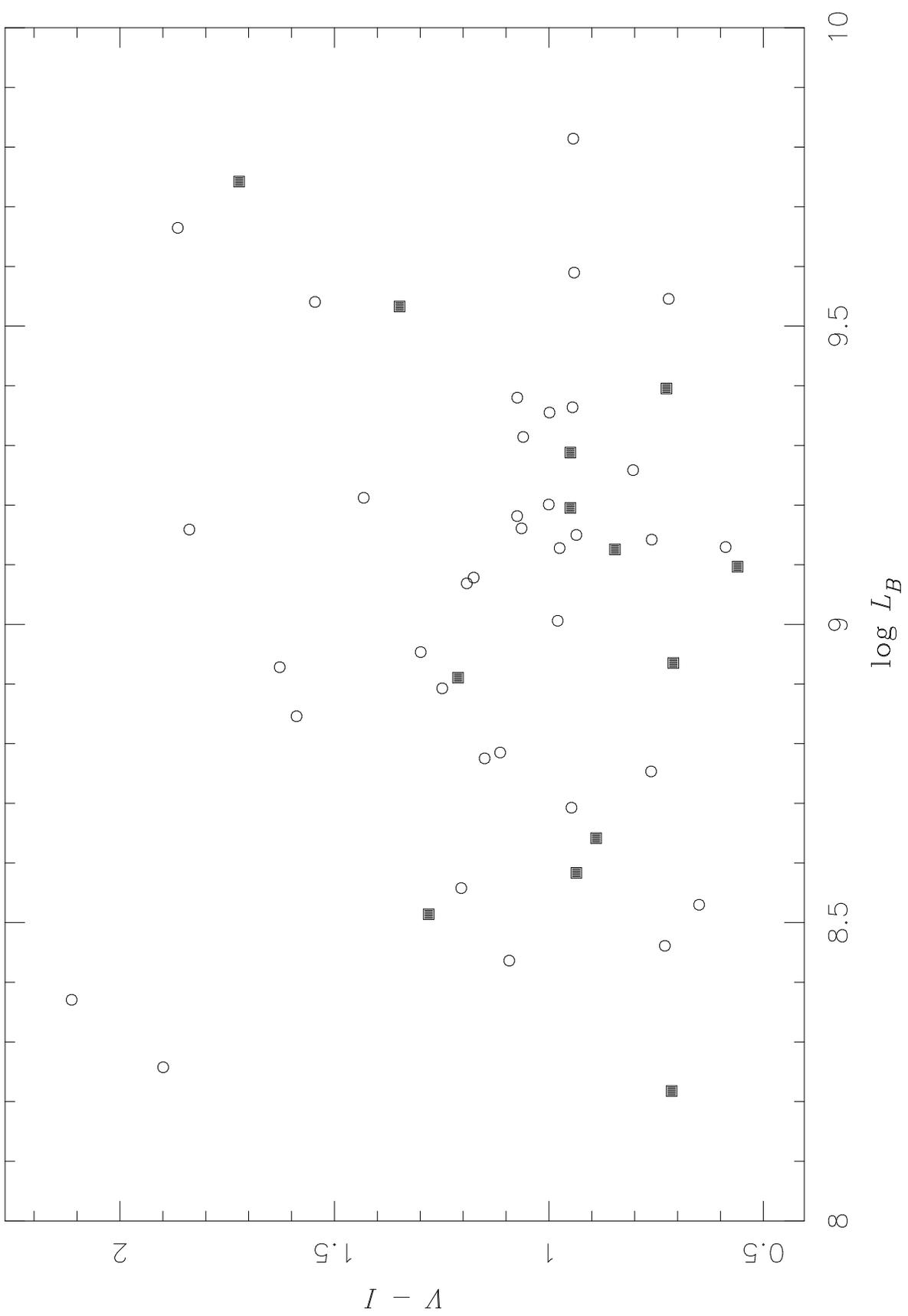

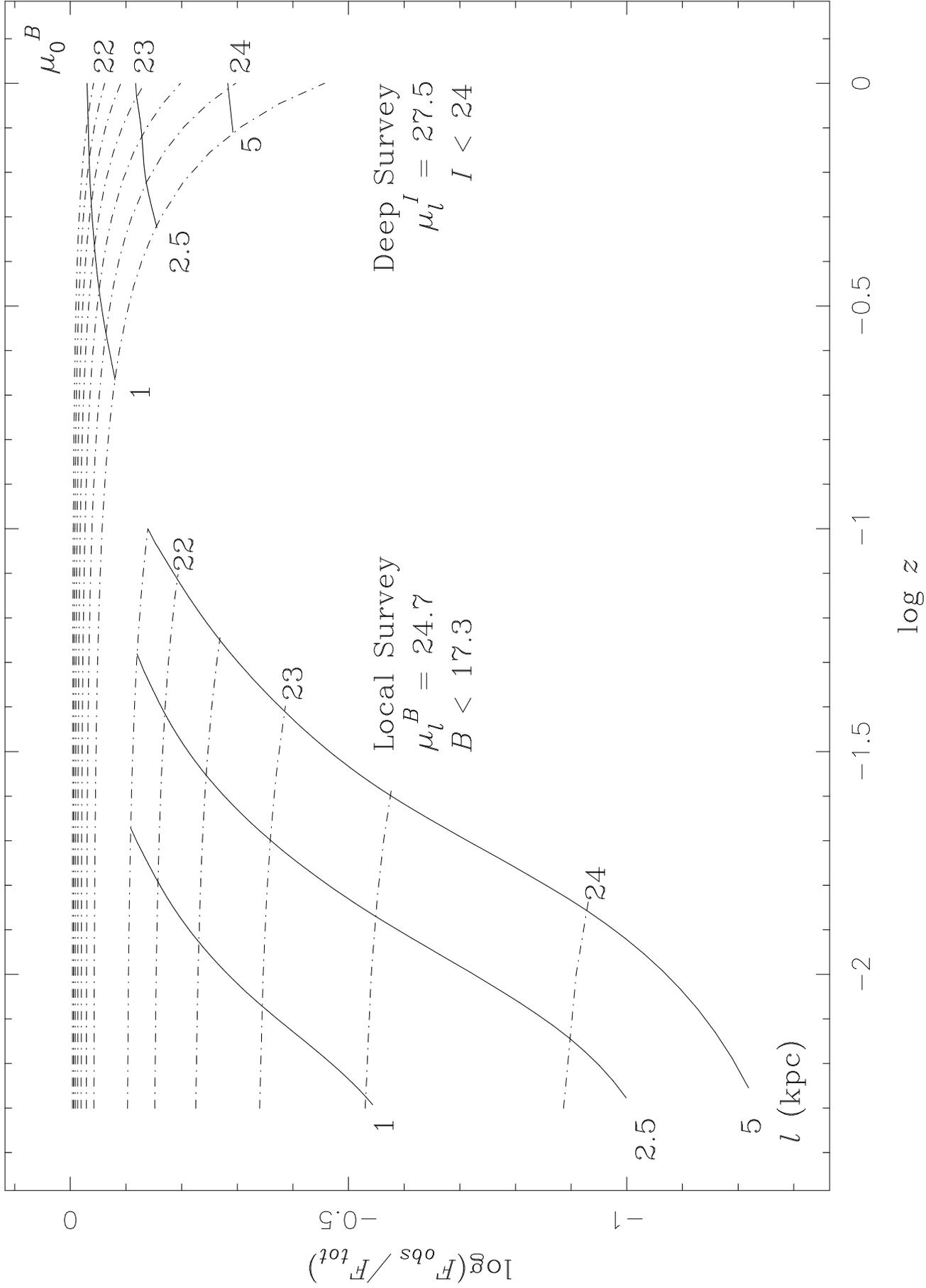

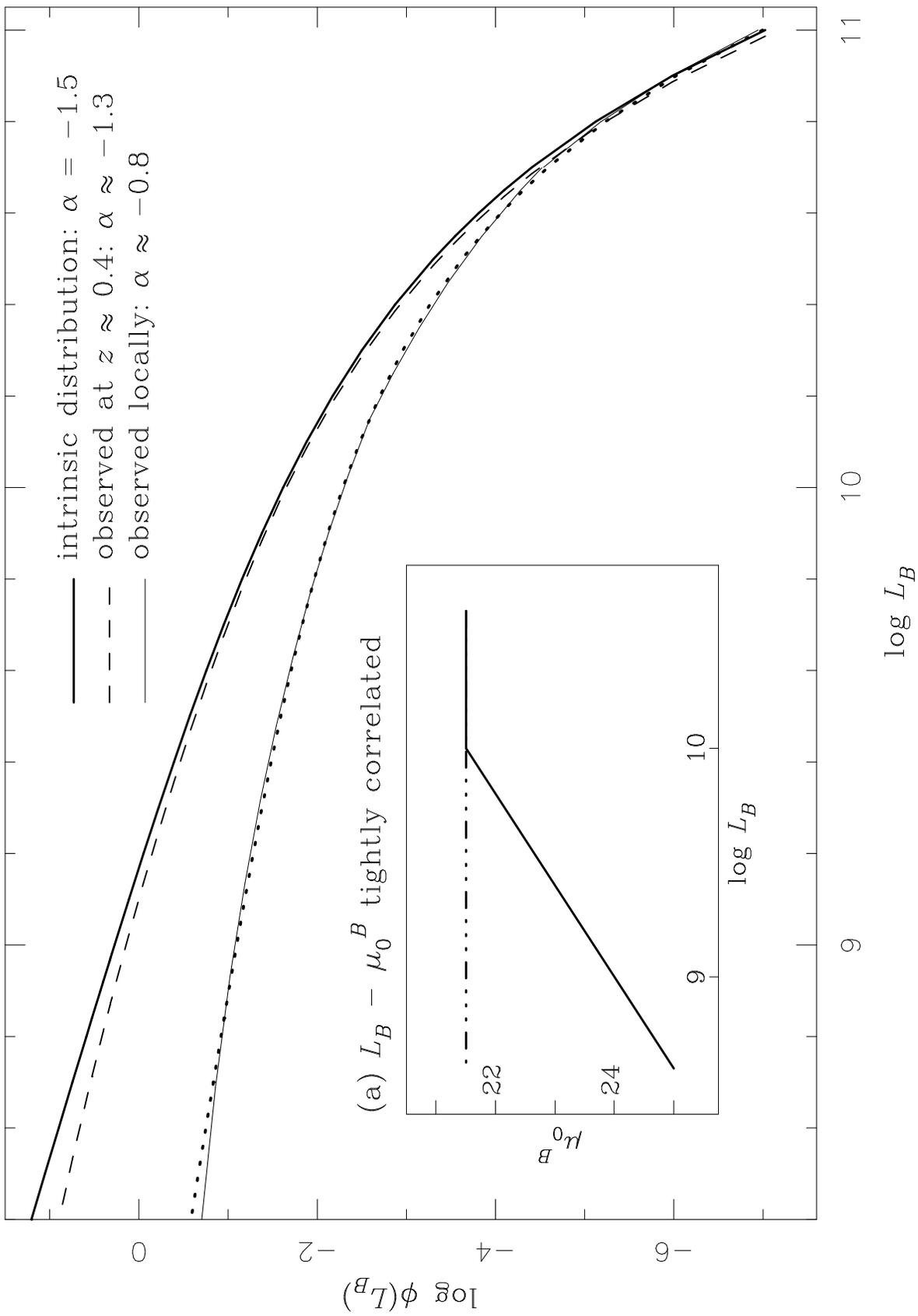

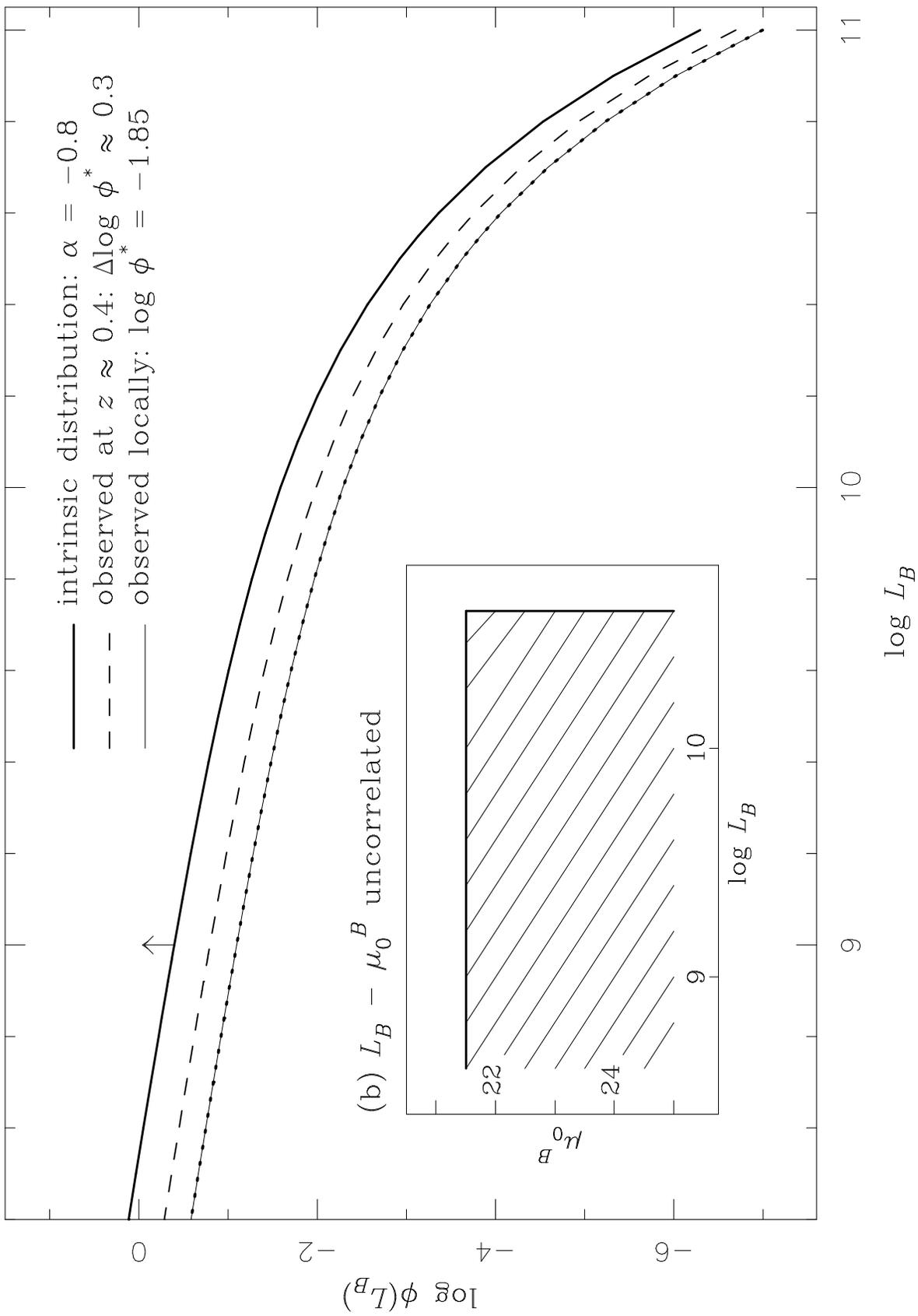